\documentclass[useAMS,usenatbib,usegraphicx]{mn2e}
\usepackage{graphicx}
\usepackage{amsmath}
\usepackage{amssymb}
\usepackage{subfigure}

\title[Two New Catalogues of Superclusters]{Two New Catalogues of Superclusters 
of Abell/ACO Galaxy Clusters out to redshift 0.15}
\author[Chow-Mart\'inez et al.]{M. Chow-Mart\'inez$^{1}$
\thanks{E-mail:marcel@astro.ugto.mx},
H. Andernach$^{1}$, C. A. Caretta$^{1}$, J. J. Trejo-Alonso$^{1}$
\\
Departamento de Astronom\'ia, DCNE, Universidad de Guanajuato, Apdo.\
Postal 144, CP 36000, Guanajuato, Gto., Mexico.\\
}
\begin{document}

\date{Accepted . Received ; in original form }
\pagerange{\pageref{firstpage}--\pageref{lastpage}} \pubyear{2011}

\maketitle
\label{firstpage}

\begin{abstract}
We present two new catalogues of superclusters of galaxies out to a redshift 
of $z\;=\;0.15$, based on the Abell/ACO cluster redshift compilation 
maintained by one of us (HA). The first of these catalogues, the all-sky 
Main SuperCluster Catalogue (MSCC), is based on only the rich (A-) Abell 
clusters, and the second one, the Southern SuperCluster Catalogue (SSCC), 
covers declinations $\delta\;<\;-17^\circ$ and includes the supplementary 
Abell S-clusters. A tunable Friends-of-Friends (FoF) algorithm was used to account for the 
cluster density decreasing with redshift and for different selection functions 
in distinct areas of the sky. We present the full list of Abell clusters 
used, together with their redshifts and supercluster memberships and including 
the isolated clusters. The SSCC contains about twice the number of 
superclusters than MSCC for $\delta\;<\;-17^\circ$, which we found to be
due to:  (1) new superclusters formed by A-clusters in their 
cores and surrounded by S-clusters (50\%), (2) new superclusters formed by 
S-clusters only (40\%), (3) redistribution of member clusters by fragmentation 
of rich (multiplicity m$\,>\,$15) superclusters (8\%), and (4) new superclusters formed by 
the connection of A-clusters through bridges of S-clusters (2\%).
Power-law fits to the cumulative supercluster multiplicity function
yield slopes of $\alpha=-2.0$ and $\alpha=-1.9$ for MSCC and SSCC 
respectively. This power-law behavior is in agreement with the findings for 
other observational samples of superclusters, but not with that of catalogues 
based on cosmological simulations.
\end{abstract}


\section{Introduction}
\label{introduction}

Superclusters are usually defined as ``clusters of clusters'' given that
mainly catalogues of galaxy clusters have been used to identify them.
\citet{devaucouleurs53} was the first to present evidence of the 
existence of a large-scale superstructure now known as the Local Supercluster. 
The superclustering as a general phenomenon was originally advocated by various
authors, beginning with \citet{abell61}, and followed by \citet{bw73}, 
\citet{hp73}, and \citet{peebles74}. 
Taking this definition of superclusters as aggregates of two or more 
galaxy clusters \citep[e.g.][]{bs84}, supercluster catalogues have been 
produced using the pseudo\,-3D distribution of Abell/ACO clusters 
\citep{abell58,aco89} by \citet{rood76}, \citet{thuan80}, \citet{bs84}, 
\citet{bb85}, \citet{west89}, \citet{zucca93}, \citet{kaku95}, and 
\citet{eina94,eina97,eina01}, among others. 
\citet{zucca93} were the first to use the total sample of Abell/ACO 
A-clusters to construct an all-sky catalogue of 69 superclusters within a 
redshift of $z\;\sim\;0.1$, with an overdensity of at least twice the 
average density, and with at least three member clusters. 
\citet{eina01} used 1663 A-clusters (64\% of them with measured spectroscopic
redshifts), out to $z\;=\;0.13$, to obtain a catalogue of 285 superclusters 
via a percolation analysis based on 
a linking length of 34\,$h^{-1}_{70}$ Mpc, assumed constant throughout the 
considered volume. The number of member clusters (commonly called multiplicity, $m$)
ranged from 2 to 35. In addition, they also presented the first catalogue of
19 superclusters based on X-ray luminous clusters detected in the ROSAT All-Sky Survey 
\citep{voges99}.  Recently \citet{chon13} published a catalogue of 195 
superclusters based on the ROSAT-ESO Flux Limited X-Ray Galaxy Cluster 
Survey (REFLEX\,II). 
Some authors have also used catalogues of individual galaxies based on 
large-scale surveys to search for superclusters, defining these as significant
density enhancement of galaxies \citep[e.g.][]{gt78,gh93,quintana95,bh98,
hanski01,eina07a,eina11a}.

Generally superclusters, being immerse in the ``cosmic web'', 
exhibit irregular or filamentary shapes
connected through bridges of galaxies and 
separated by extensive ``void'' regions where almost no galaxies are 
found. Properties of the large-scale structure (determined using both observational 
data and N-body cosmological simulations) have been discussed, e.g.\ by
\citet{jaan98}, \citet{gs02}, \citet{kolok02}, \citet{wray06},
\citet{eina07b,eina07c,eina07d,eina11b}, \citet{costaduarte11},
\citet{lupare11}, \citet{sousbie11a}, and \citet{sousbie11b}.

Currently there is no evidence that superclusters have reached virialization, 
since the sizes of these structures range 
from a few Mpc to $\sim$150\,$h^{-1}_{70}$ Mpc, and the crossing time for
a member cluster within the system exceeds the age of the Universe 
\citep[e.g.][]{oort83,gs02}. This implies that these systems still preserve 
a memory of their dynamical history, which makes them worth studying.
Furthermore, since they constitute the environment of a considerable 
fraction of the clusters, groups and galaxies themselves, by comparing 
their properties with those in lower-density environments we can study 
the effect of this environment on the evolution of such systems.
Galaxy clusters are the most massive systems 
in the Universe that have condensed out of the Hubble flow, and their evolution 
is still a matter of vital discussion. Finally, superclusters represent regions 
of significant density enhancement which cause distortions in the local 
gravitational field that are sometimes noticeable via the bulk motions towards 
them \citep[e.g. the case of the Great Attractor,][]{lyndenbell88}. 
They also provide information on the mass distribution in intercluster space, 
e.g.\ by measuring their imprint on the CMB via the Sunyaev-Zeldovich effect 
\citep[e.g.][]{planck13}. All these features allow one to constrain and refine 
cosmological models of the Universe.

In the present work we describe two new catalogues of superclusters based
on the update of late 2012 of the Abell/ACO cluster redshift compilation
by one of us \citep[for a description of its content cf.][]{ander05},
restricting ourselves to clusters with redshift $z\;\leq\;0.15$.  For the
first catalogue we consider the distribution of Abell/ACO A-clusters
(rich clusters) all over the sky, while the second one is a southern
sky catalogue ($\delta\;<\;-17^\circ$) based on both A-clusters and the
supplementary S-clusters (typically poorer than the A-clusters). For
the construction of both catalogues we tuned the linking length as a
function of redshift and position on the sky, thus allowing for both
undersampling at higher redshifts and deeper redshift observations
over certain areas of the sky, like those covered by the Sloan Digital
Sky Survey \citep{abazajian09}. The effect of the inclusion of the
S-clusters in the large-scale structure is studied by comparing the
two catalogues. We also discuss the cumulative distribution of multiplicities 
(number of member clusters) for the superclusters in the Local Universe, 
based on both observational and simulated data.

The paper is organized as follows. In section \ref{sec:data} we describe 
our Abell/ACO cluster sample; in section \ref{sec:analysis} we discuss and 
explain the percolation analysis we applied; section \ref{sec:sc} explains the 
generated catalogues and compares them with that of \citet{eina01}; 
in section \ref{sec:prop_sc} 
we present some properties of the large-scale structure by intercomparing
our catalogues, and by comparing their multiplicity functions
with the ones obtained for different supercluster samples, including 
supercluster catalogues generated from distributions of equivalent dark 
matter haloes in cosmological simulations. Section~\ref{sec:conclusion} presents
our conclusions and summary.

Throughout this paper we assume the following cosmological parameters: 
$H_0\,=\,70\,h_{70}$ km s$^{-1}$ Mpc$^{-1}$, $\Omega_M\,=\,0.3$, and 
$\Omega_\Lambda\,=\,0.7$.

\section{Data}
\label{sec:data}

The Abell/ACO cluster redshift compilation used here \citep[e.g.][]{ander05} 
is a collection of individual radial velocities for potential 
member galaxies of such clusters. The compilation has been updated since 1989 
\citep{ander91, ander95, ander98} by monitoring redshift data of galaxies from 
the published literature and a few unpublished references. 
The compilation, as of late 2012, contains redshifts for about 130,000 
individual cluster galaxies in 3930 different Abell clusters. 
Whenever a cluster shows more than one concentration along the line of sight, 
separated in redshift by more than $\sim$1\,500\:km\,s$^{-1}$, we
register these as components A, B, C,... in increasing order of redshift.
In many cases, the component with the dominant number of redshifts is clearly 
the one recognized by Abell/ACO, while in other cases these components
may be similarly populated, and the identification is not straightforward.
In what follows, we shall refer to these as \textsl{line-of-sight 
components} of a cluster, and we include all the detected ones in our analysis 
out to $z=0.15$.

With the advent of large-scale multi-object spectroscopic surveys, 
like the Two-Degree Field Galaxy Redshift Survey \citep[2dFGRS,][]{colles01},
the Six-Degree Field Galaxy Redshift Survey \citep[6dFGRS,][]{jones04}, and 
the Sloan Digital Sky Survey \citep[e.g. SDSS-DR7,][]{abazajian09}, 
the source of redshifts for this compilation has gradually changed from surveys 
of individual clusters to large-area redshifts surveys, resulting in more 
redshifts for cluster members from fewer references per year. We include 
a number of new spectroscopic redshifts obtained by us for galaxies in 121 
clusters. These will be published in a separate paper. Furthermore, for clusters 
with no spectroscopic redshift as yet, the compilation also includes estimated 
redshifts, kindly provided by M.\ West (2007, private communication) and based 
on the relation proposed by \citet{pw92}.

\begin{figure}
\includegraphics[width=8cm]{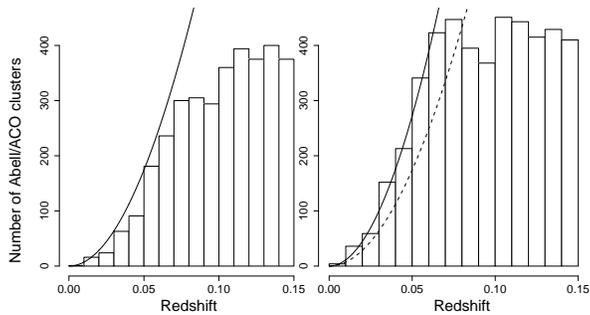}
  \caption{Redshift distribution for the A-clusters (left panel) and A+S clusters 
(right panel). The solid lines are the expected number of clusters in spherical 
shells of $\Delta z=0.01$ using a constant space density of clusters 
corresponding to the most completely surveyed redshifts for each distribution, 
namely $0.03<z<0.07$ for the all-sky distribution of A-clusters, and $z\;<\;0.05$ 
for the A+S clusters in the southern ($\delta<-17^\circ$) sample.
These values are: 7.4$\times10^{-6}$ $h^{3}_{70}$ Mpc$^{-3}$ for only the 
A-clusters and 2.8$\times10^{-5}$ $h^{3}_{70}$ Mpc$^{-3}$ for 
A- and S-clusters together in the south.
The first value is comparable to the one found by \citet{eina97} 
(8.9$\times10^{-6}$ $h^{3}_{70}$ Mpc$^{-3}$) for Abell/ACO clusters.
The dashed line in the right panel indicates the same function as the 
solid line in the left panel. These curves grow according 
to $r^2$ where $r=r(z)$ is the comoving distance as described by
\citet{hogg00}.
}
\label{figure:sel}
\end{figure}

The current compilation includes all 4076 A- and 1174 S-clusters. For 1011 of 
the A-clusters we list a total of 2511 line-of-sight components,
resulting in altogether 5576 clusters/components, while 230 S-clusters
are split into 551 components, resulting in a total of 1495 S-clusters/components.
From these we selected the ones with redshift below 0.15 as the basis 
for our search for superclusters. 
These were 3410 A-clusters/components (92\% of them having measured 
spectroscopic redshifts for at least one member, and 76\% for at least three 
galaxies), and 1168 S-clusters/components (91\% with at least 
one galaxy with measured spectroscopic redshift, and 69\% with at least three). 
Beyond this redshift the sample becomes noticeably more incomplete.
The redshift distribution for both samples is shown in Figure~\ref{figure:sel}. 
The S-clusters represent an important contribution in the southern sky, and 
they complement the sample toward lower richness clusters in this region 
at lower redshifts. 

\begin{figure}
\includegraphics[width=8cm]{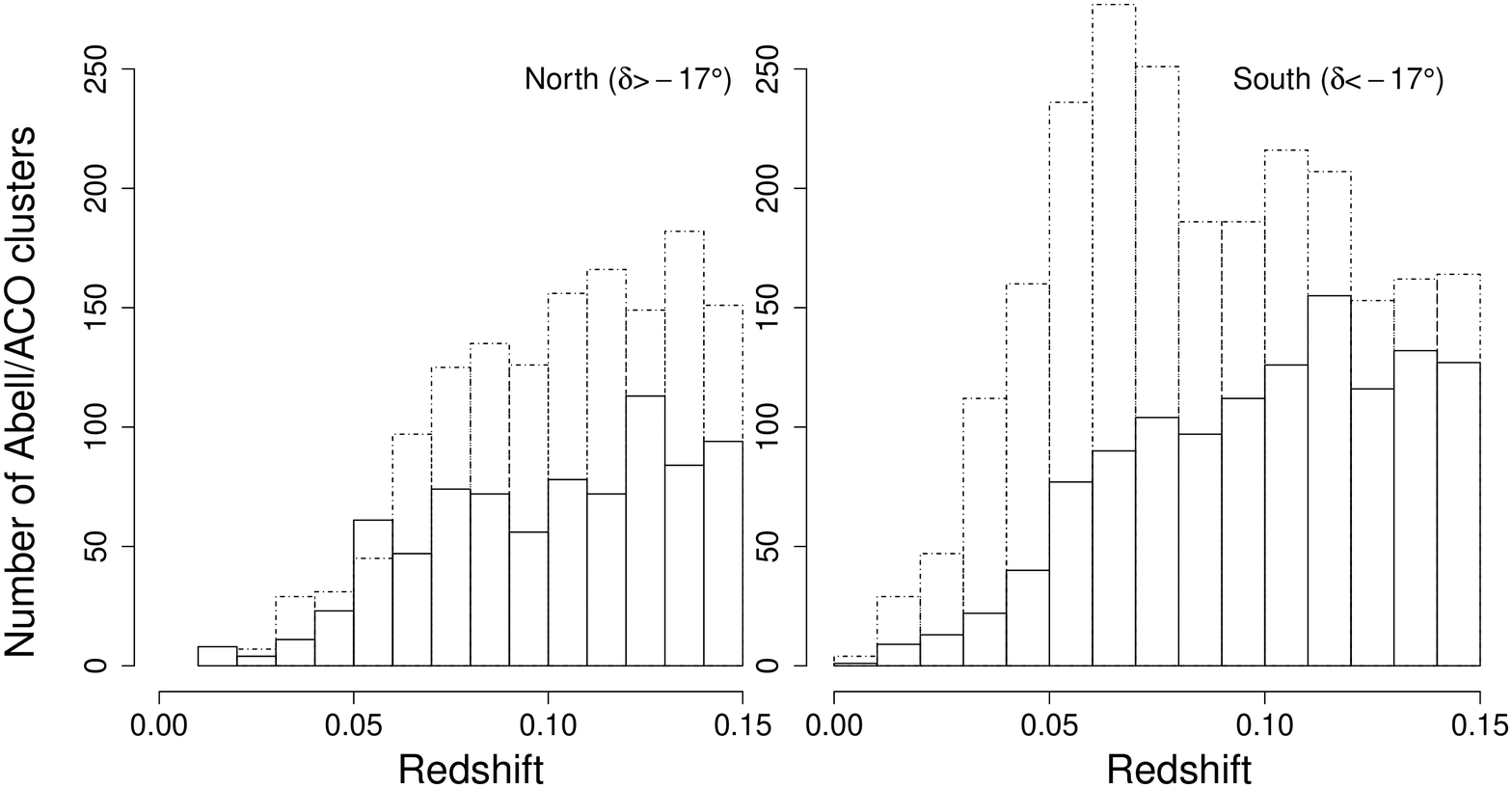}
  \caption{Redshift distribution for the northern and southern Abell/ACO 
clusters. For the northern subsample, the solid histogram is for the northern 
sky without the SDSS region and the dashed histogram is the one for 
the SDSS region. For the southern subsample, the solid histogram is the 
distribution for A clusters and the dashed histogram is the one for A+S clusters.
}
\label{figure:zdis}
\end{figure}

In Figure~\ref{figure:zdis} we divided the entire sample in subsamples
according to their position on the sky, in order to identify systematic 
effects in their global distribution. In the northern sample the main effect 
is the presence of the SDSS redshift data: where these data exist the clusters 
are more completely spectroscopically sampled such that more clusters enter 
in the sample at relatively high redshifts ($z\gtrsim0.06$). On the other 
hand, the S-clusters affect the southern sample especially at lower redshifts 
($z\lesssim0.08$) such that the first peak in redshift in the second panel 
of both Figs.~\ref{figure:sel} and \ref{figure:zdis} is dominated by them. 
In the redshift range considered here, the S-clusters are almost exclusively 
poorer systems than the A-clusters. Thus, we decided to divide our sample 
into four, rather than two, subsamples, namely two northern subsamples, one 
for the SDSS region and another for the rest of the northern sky, and another 
two southern subsamples ($\delta<-17^\circ$), one including both A- and 
S-clusters and the other only the A-clusters.

\section{Percolation Analysis}
\label{sec:analysis}
The Friends-of-Friends algorithm \citep[FoF, e.g.][]{turner76,zeld82,hg82,eina84}
is frequently used to link points when we do not know \textit{a priori}
how they are distributed.  In a virialized regime, such as we expect
for at least the cores of the galaxy clusters, the linking length
is determined by assuming a spherical gravitational collapse model 
\citep[e.g.][]{caretta08}.
However, since the superclusters are probably not in an equilibrium state, we can
not derive the ideal linking length in this way.  

In the percolation analysis, the method usually used for identifying
superclusters of galaxies, the linking length of the FoF is chosen 
such that it maximizes the number of systems formed.
The result of the percolation analysis depends strongly on the linking
length ($\ell$).  For smaller linking lengths
the FoF yields a smaller number of only the densest systems, while for
larger linking lengths these systems start to connect among themselves,
lowering the total number of systems until a complete percolation of the
sample is reached \citep[e.g.][]{eina84}. For this reason, in the identification of
superclusters of galaxies, the linking length of the FoF is often chosen
such that it maximizes the number of systems formed in the percolation 
process. We shall call this the \textit{critical linking length}, $\ell_c$.
The percolation analysis relies on the fact that the mean
density is correctly represented (i.e.\ the completeness of the sample) 
in the considered volume. In an ideal (complete and homogeneous) sample, the 
critical linking length obtained this way should be the same throughout 
the entire volume. Since the distribution of the clusters in our sample is 
clearly less complete at higher redshifts (see Figures~\ref{figure:sel} and 
\ref{figure:zdis}), and also differs from one subsample to the other, the 
choice of a single critical linking length for a supercluster search based 
on Abell/ACO clusters is certainly not the best solution. 
Therefore, we used a \textsl{tunable} value of $\ell_c$, which is allowed
to vary with the selection function and thus corrects for completeness 
variations in both redshift and position on the sky. This selection 
function (the fraction of clusters missed as a function of the redshift, due to 
an inhomogeneous sampling and the Malmquist bias) is measured directly by 
estimating the value of $\ell_c$ in the four different subsamples and 
in different redshift ranges.

\begin{figure}
\centering
\includegraphics[width=8cm]{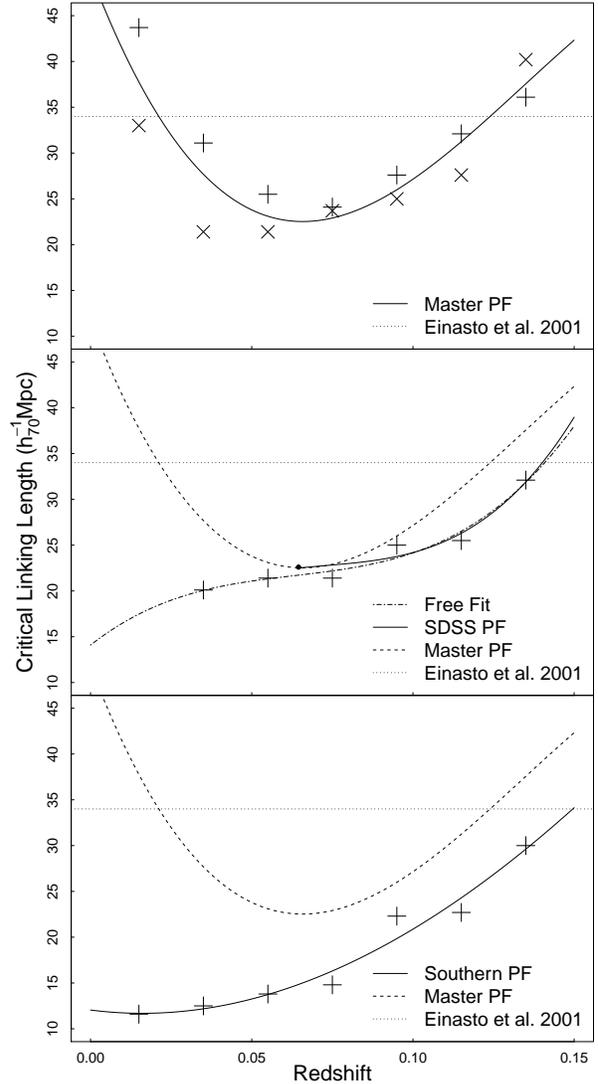}
\caption{\textsl{Percolation functions} (PFs) for the considered samples. 
Top panel: PF for the Northern region (excluding the SDSS) 
and the Southern region (excluding S-clusters), the former marked 
with ``+'' signs and the latter one with ``$\times$'' signs; the fit 
is what we call the \textsl{Master PF}. 
Middle panel: PF for the SDSS region, the dashed line is the 
Master PF, the dot-dashed line is a free fitting for the SDSS region, and the
solid line is the curve we call the \textsl{SDSS PF}, in which 
the SDSS PF is forced to follow the Master PF for $z\leq0.06$. 
Bottom panel: PF for the complete southern sample (including 
S-clusters), where the Master PF is again represented by the dashed line 
and the Southern PF by the solid line. For comparison, the linking length
used by \citealt{eina01} is indicated with a dotted line.
}
\label{figure:per}
\end{figure}

To this end we divided each subsample into seven spherical shells with 
a width of $\Delta z=0.03$, (allowing an overlap between them of 
$\delta z=0.01$), inside which the density is not expected to vary
significantly. Thus, the first bin ranges from $z=0.00$ to $z=0.03$, the 
second one from $z=0.02$ to $z=0.05$, and so on. 
We obtained the value of $\ell_c$ for each redshift ``shell'' and constructed 
the so-called ``Percolation Function'' (PF), i.e.\ the value of  $\ell_c$ 
as function of the redshift. Percolation functions for each
subsample are shown in Figure~\ref{figure:per}. Three different PF
were chosen. The first of these is the combination of the northern sample 
(excluding the SDSS region) and the southern sample (excluding the
S-clusters), which we call the \textsl{Master~PF}. As expected, these two 
subsamples show a similar behavior. 
The second PF is that for the SDSS region, which we call the \textsl{SDSS~PF}. 
Since it is poorly sampled at lower redshifts ($z<0.06$), we first 
freely fitted the data and, then, forced it to coincide with the 
Master~PF at $z=0.06$, taking the Master~PF as the SDSS~PF for redshifts
lower than this. The third PF is that for the southern subsample including the
S-clusters, which we call the \textsl{Southern~PF}. This latter one
presents smaller critical linking lengths, $\ell_c$ than the Master~PF at every redshift, 
and especially so at the lowest redshifts, obviously due to the higher space 
density of clusters when S-clusters are included.

The redshift range in which the PFs reach their lowest value may be
interpreted as the one of highest completeness, and {\it vice-versa}:
in an undersampled region, the value of  $\ell_c$ is larger. 
Thus, employing these PFs in our FoF compensates for the loss of objects due to selection
effects. According to these curves, the best sampled region (excluding
the contribution of SDSS and S-clusters) is near $z\sim0.06$, where the
Master~PF reaches its minimum. At lower redshifts the sample suffers
a deficiency due to the exclusion of clusters at $z\leq0.02$, as imposed
by the authors of the Abell/ACO catalogues. At higher redshifts the
undersampling is due to limiting sensitivity of the sky surveys used to
find the clusters and incompleteness in the spectroscopic sampling. 
The SDSS~PF has smaller values than the Master~PF at higher 
redshifts due to the more complete sampling of the SDSS.  As a consequence 
of the inclusion of the S-clusters, the Southern~PF is always smaller than 
the Master~PF, especially at lower redshifts. 
Near the minimum of the Master~PF (i.e.\ where the sampling is best, near $z=0.06$) 
we find  $\ell_c\sim23\,h^{-1}_{70}$ Mpc while for 
lower redshifts the Southern~PF reaches a minimum that corresponds to a
$\ell_c\sim12\,h^{-1}_{70}$ Mpc.

\section{Superclusters of Abell/ACO galaxy clusters}
\label{sec:sc}
\subsection{The Catalogues}
\label{sec:cat}

The Master and SDSS PFs were used to produce the {\it Main SuperCluster
Catalogue} (MSCC) based on the all-sky sample of A-clusters. The MSCC
contains 601 superclusters with multiplicities between 2 and 42. 
The {\it Southern SuperCluster Catalogue} (SSCC) was produced using the 
Southern PF, and contains 423 superclusters with multiplicities between 2 and
38. For the rest of this paper we shall refer to individual superclusters
with their MSCC and SSCC acronyms, followed by their sequence number 
(see Tables~\ref{table:mscc} and \ref{table:sscc}).

For the MSCC, 1152 clusters (34\% of the total number of clusters)
turn out to be isolated, while this number is 870 (36\%) for the SSCC.
The fractions of superclusters with multiplicity $m=2$
(pairs of clusters) are 48\% and 52\% (containing 17\% and 19\% of the input 
clusters), in the MSCC and SSCC, respectively. The richest superclusters,
with $m>10$, represent 4\% in both MSCC and SSCC, whose member clusters 
constitute 49\%, and 45\% of the input clusters for MSCC and SSCC, respectively.
We also flagged as \textsl{supercluster candidates} the ones which 
would not be found by our FoF in a test catalogue that excludes the clusters
with estimated redshifts. In our Abell/ACO sample limited to $z\leq0.15$,
out of 4578 clusters (including A- and S-clusters as well as their 
line-of-sight components) there are 358 with only estimated redshifts,
identified with an ``e'' following their names in Tables~\ref{table:mscc} and \ref{table:sscc}.
Using the above definition, 37 (6\%) of the MSCC and 38 (9\%) of the SSCC 
superclusters are candidates, identified in the catalogues with a letter 
``c'' following the supercluster sequence number.

\begin{table*}
  \centering
\caption{First ten entries in MSCC: (1) MSCC number; (2) SCL identification; 
(3) multiplicity, $m$; (4,5) right ascension and declination
(J2000) in decimal degrees; (6) redshift, $z$; 
(7) distance, in $h^{-1}_{70}$ Mpc, of the pair of clusters 
with the maximum separation in each supercluster;
(8) list of member clusters of the supercluster. 
The full table is available online.}
\begin{tabular}{rrrrrrr p{10cm}}
  \hline
  \multicolumn{1}{c}{MSCC}                &
  \multicolumn{1}{c}{SCL}                 &
  \multicolumn{1}{c}{$m$}                 &
  \multicolumn{1}{c}{RA$_J$}   &
  \multicolumn{1}{c}{Dec$_J$}  &
  \multicolumn{1}{c}{$z_{SC}$}            &
  \multicolumn{1}{c}{d$_{\sf max}$}  &
  \multicolumn{1}{c}{Abell/ACO member clusters}   \\
\hline
  1~  & $-$ & 9  & 0.77 & $-$26.72 & 0.064  & 50.6 & A0014, A0020, A2683A, A2716, A2726A, A2734, A4038C, A4049B, A4053B  \\
  2~  & 3   & 2  & 1.09 & $+$09.77 & 0.098  & 20.2 & A2694, A2706  \\
  3~  & $-$ & 4  & 1.20 & $+$16.05 & 0.119  & 41.5 & A0001, A2688A, A2703, A2705  \\
  4~  & 3   & 4  & 1.26 & $+$02.89 & 0.098  & 33.4 & A0003, A2696B, A2698, A2700A  \\
  5~  & $-$ & 2  & 1.31 & $-$15.32 & 0.102  & 20.3 & A2699, A2710A  \\
  6~  & $-$ & 4  & 1.58 & $-$64.24 & 0.116  & 37.7 & A2732, A2740A, A2760, A4028  \\
  7~  & 220 & 3  & 2.03 & $-$36.93 & 0.049  & 28.5 & A2717A, A2771A, A4059  \\
  8~  & 5,9 & 10 & 2.09 & $-$35.26 & 0.117  & 79.4 & A2715B, A2721A, A2721B, A2724, A2730, A2749B, A2767, A2772, A4035A, A4074A  \\
  9~  & $-$ & 2  & 2.35 & $-$10.57 & 0.110  & 16.2 & A0008A, A2709B  \\
  10~ & $-$ & 3  & 3.06 & $-$34.91 & 0.096  & 22.4 & A2715A, A2749A, A2755  \\
\hline\end{tabular}
\label{table:mscc}
\end{table*}
\begin{table*}
  \centering
\caption{First ten entries in SSCC: (1) SSCC number; (2) SCL identification; 
(3) multiplicity, $m$; (4,5) right ascension and declination
(J2000) in decimal degrees; (6) redshift, $z$; 
(7) distance, in $h^{-1}_{70}$ Mpc, of the pair of clusters
with the maximum separation in each supercluster;
(8) list of member clusters of the supercluster. 
The full table is available online.}
\begin{tabular}{rrrrrrr p{10cm}}
  \hline
  \multicolumn{1}{c}{SSCC}                &
  \multicolumn{1}{c}{SCL}                 &
  \multicolumn{1}{c}{$m$}                 &
  \multicolumn{1}{c}{RA$_J$}   &
  \multicolumn{1}{c}{Dec$_J$}  &
  \multicolumn{1}{c}{$z_{SC}$}                 &
  \multicolumn{1}{c}{d$_{\sf max}$}  &
  \multicolumn{1}{c}{Abell/ACO member clusters}   \\
\hline
  1~ & $-$  & 2  & 0.20 & $-$23.73 & 0.097 & 18.8 & A2681, A2719     \\
  2~ &  5   & 11 & 0.68 & $-$34.92 & 0.116 & 80.0 & A2715B, A2721A, A2721B, A2730, A2749B, A4035A, A4074A, S0012B, S1161C, S1161D, S1170B     \\
  3~ & $-$  & 2  & 0.80 & $-$44.78 & 0.038 & 10.9 & S0005, S1173     \\
  4~ & $-$  & 3  & 1.23 & $-$38.67 & 0.103 & 24.2 & A4068, S0017e, S1172C     \\
  5~ & $-$  & 2  & 2.61 & $-$30.93 & 0.070 & 14.0 & A2751A, S0006D     \\
  6~ & $-$  & 2  & 3.38 & $-$29.83 & 0.123 & 24.7 & A2759B, S0010     \\
  7c & $-$  & 2  & 3.91 & $-$65.43 & 0.148 & 24.2 & A2737e, A2761e     \\
  8~ & $-$  & 5  & 4.23 & $-$64.81 & 0.114 & 31.7 & A2732, A2740A, A2760, S0018e, S0057e     \\
  9~ & $-$  & 2  & 4.24 & $-$35.03 & 0.095 & 4.2  & A2749A, A2755     \\
 10~ & $-$  & 2  & 4.34 & $-$23.26 & 0.066 & 11.8 & A0014, A0020     \\
\hline\end{tabular}
\label{table:sscc}
\end{table*}

We also include in Tables~\ref{table:mscc} and \ref{table:sscc} the
supercluster ID of \citet{eina01} for some specific cases (187 for MSCC and 
117 for SSCC) where a certain minimum fraction of member clusters (explained
below) in that catalogue is linked to the respective supercluster here. From now
on, we shall denote \citet{eina01} superclusters ID with the prefix SCL,
as chosen by these authors.  
Tables~\ref{table:mscc} and \ref{table:sscc} only include the 
cross-identifications when the association was sufficiently clear, with
minimum ambiguities by fragmentations or ``noise'' from subcomponents (divisions of the
superclusters due to their member clusters having line-of-sight components).
We assumed an entry in SCL to be the same as one in MSCC/SSCC when the 
number of coincident member clusters exceeded a certain fraction 
or when the main member cluster of well-known superclusters
(e.g. those listed in Section~\ref{sec:individual}) was found to be the same.
The minimum fraction of member clusters required for the match depended on 
the multiplicity in MSCC: for pairs, we required both clusters to coincide;
for triple and quadruple systems we required at least two equal member 
clusters. For higher multiplicities ($m\geq\;5$) we accepted ``fragments''
(i.e.\ a subsample of member clusters) of the original superclusters listed 
in SCL, and, as we said, in some cases we allowed fractions lower than $50\%$ 
when the main member clusters were present in some particular MSCC. 
Where more than one entry in SCL match the same MSCC/SSCC, all SCL ID
of these contributers are included (e.g. MSCC~8 in Table~\ref{table:mscc}),
in order of decreasing fraction of coincident member clusters
(cases for which only one coincident cluster was found are not listed).
For SSCC, as we will discuss in Section~\ref{sec:matchsscc}, this 
fragmentation is stronger, so we accepted an identification with SCL when we had at 
least two coincidences of original member clusters for the fragments of 
richest superclusters. This was done because fragments usually are 
accompanied by S-clusters in SSCC.
In Section~\ref{sec:matchscl} we explain this matching procedure in more 
detail, including our treatment of clusters with line-of-sight components. 
There we also present more complete statistics accounting for more complex 
cases. 

The MSCC and SSCC catalogues (Tables~\ref{table:mscc} and \ref{table:sscc}) 
contain the following columns: (1) the supercluster ID in the respective 
catalogue followed by the letter ``c'' if the supercluster is considered a candidate; 
(2) the matching SCL from \citet{eina01} where applicable; two or more, at most 
four, SCL ID's appear if we found these SCL's identifyable with
the same MSCC or SSCC;
(3) the multiplicity $m$, i.e.\ the number of member clusters; 
(4) right ascension and (5) declination (for equinox J2000, in decimal degrees),
obtained as the simple arithmetic mean of the member cluster positions; 
(6) the arithmetic mean redshift $z_{SC}$ of the member clusters, including
clusters with estimated redshift; 
(7) the 3D-distance between the most separate pair of member clusters
in each supercluster, in $h_{70}^{-1}$ Mpc, as an estimation of the supercluster
size;
(8) the list of the member clusters, 
appended by an ``e'' if that cluster had only an estimated redshift. 
Letters A, B, C, ...  appended to the cluster name indicate a line-of-sight 
component in order of increasing redshift.

Table~\ref{table:cluster} lists all clusters used in our analysis, both
isolated ones and supercluster members, including all line-of-sight
components within the considered volume of $z\le0.15$.
Again, where the name of the cluster is followed by the flag ``e'' in the first 
column of Table~\ref{table:cluster}, the listed redshift is estimated.
The third and fourth columns indicate the membership in MSCC or SSCC, where
isolated clusters are marked with ``iso''.
For line-of-sight components of clusters (labelled A, B,..., etc.), we list
the percentage of galaxies with measured redshift that contribute to the 
line-of-sight component.  Since not all line-of-sight components are inside the 
examined volume ($z\leq0.15$), in a few cases the sum of the fraction is lower 
than 100\%.

\begin{figure*}
\centering\includegraphics[width=0.98\textwidth]{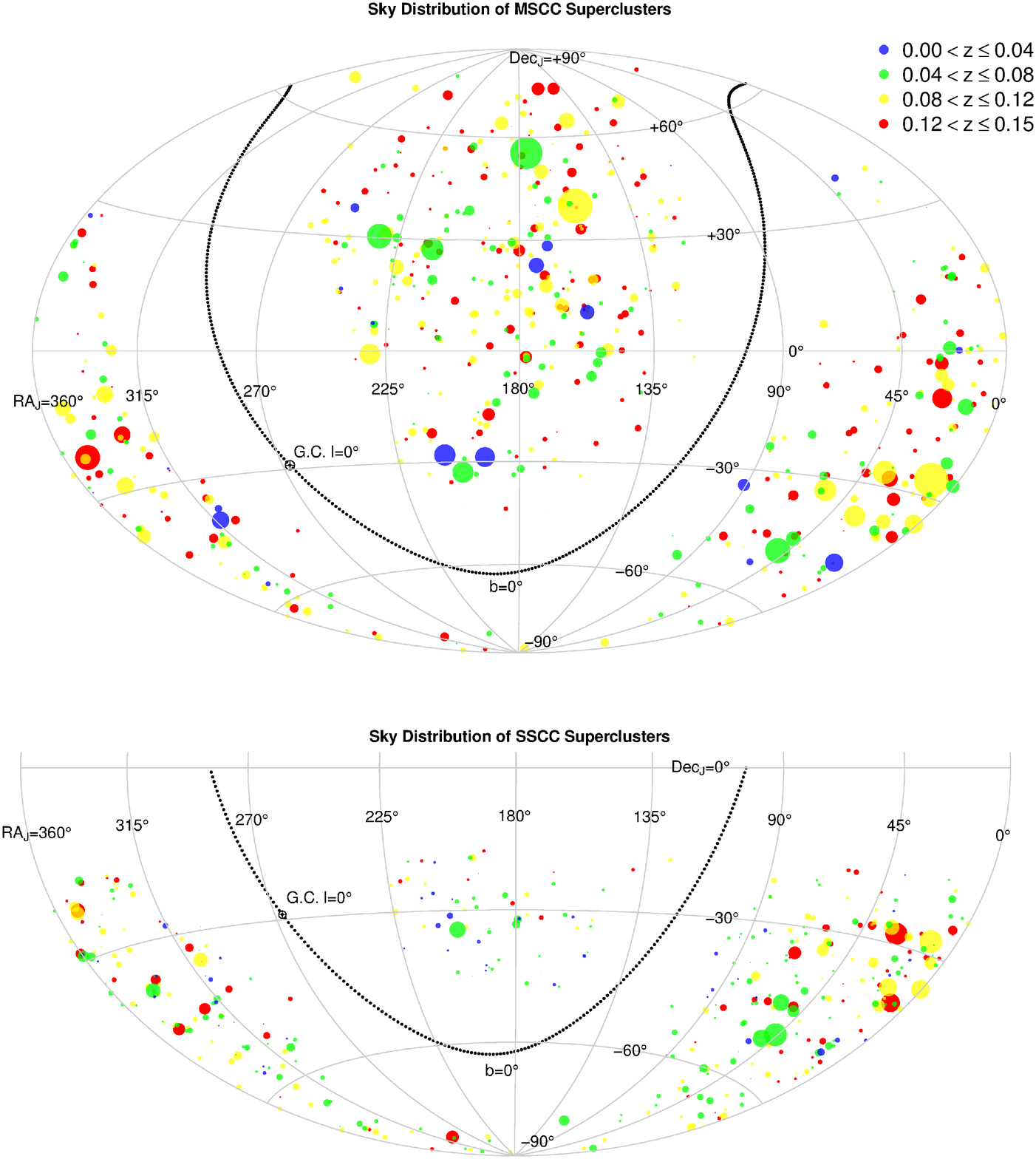}
\caption{Sky distribution in an Aitoff projection of equatorial coordinates 
of superclusters in MSCC (top panel), and SSCC (bottom panel). The superclusters
are colored according to their location in four redshift intervals (see
legend at top right). The symbol size is proportional to the separation d$_{\sf max}$,
of member cluster pairs in each supercluster (cf.\ Table~\ref{table:mscc}), and the
ratio of symbol size to d$_{\sf max}$ decreases slightly with increasing redshift interval.
In both projections the black dotted line indicates the Galactic plane ($b=0^\circ$)
and ``G.C.'' denotes the Galactic centre.
}
\label{figure:mappro}
\end{figure*}

The projected sky distributions of both catalogues are displayed 
in Figure~\ref{figure:mappro}.

\subsection{Comparison with Previous Supercluster Catalogues}
\label{sec:matchscl}

A comparison of the MSCC and SCL catalogues should give an idea of the improvements
achieved over the last 15 years, mainly due to (a) the inclusion of new redshifts
from the recent literature, (b) the extension to a higher redshift limit, 
and (c) the inclusion of all redshift components for the individual clusters,
when they are present.  \citet{eina01} found 285 superclusters based on a 
critical linking length of $\ell_c=34\,h^{-1}_{70}$\,Mpc, assumed to be constant
throughout the considered volume out to $z_{lim}=0.13$.
Rather than all line-of-sight components, these authors used only the component 
with the largest number of spectroscopic redshifts, assuming this one to be 
the ``correct'' Abell/ACO cluster. The cluster sample used by \citet{eina01} 
consisted of 1663 A-clusters, 64\% of which had measured redshifts. 
Of these, 1163 clusters (70\%) were found to be members of pairs or 
superclusters, about 5\% more than what we found for MSCC and SSCC.

To compare our MSCC with the SCL catalogue we performed a cluster-by-cluster 
cross-matching. In the case of clusters with line-of-sight components,
only the dominant component was taken into account for the comparison,
defining this as the component with the largest number of spectroscopically 
measured galaxies \citep[in an attempt to match the definition used 
by][]{eina01}.

Based on the above, of the total number of clusters (1159) which are members 
of any of the 285 ``SCL'' superclusters, 880 (76\%) were located within 356 
superclusters of MSCC, while another 279 (24\%) were found to be isolated in 
MSCC. Moreover, the remaining 245 MSCC superclusters, which are
new compared to SCL, can be divided into: 112 (46\%) which are beyond
the redshift limit of SCL ($z_{lim}=0.13$) and another 14 (6\%) which are
close to that limit, such that they have members on both sides of that limit; 
another 47 (19\%) resulted to be actually ``components'' in the line of 
sight of superclusters, formed mostly by part of the members of the ``dominant'' 
supercluster; 35 (14\%) are superclusters comprised by line-of-sight components 
accompanying clusters that were considered isolated in SCL; and finally, 37 
(15\%) represent completely new superclusters within the volume considered 
by \citet[][]{eina01} \footnote{One of these cases is the pair MSCC~164, 
formed by the clusters A0569N and A0569S \citep[e.g.][]{beers91}.}.

Of the 255 superclusters of SCL whose clusters have a match in
some MSCC supercluster, 108 (42\%) had a one-to-one match
in MSCC, while 106 (42\%) were ``fragmented''
(member clusters were linked into different superclusters in MSCC). 
Another 41 (16\%) SCL were redistributed in more complex ways; their member
clusters are part of one unique MSCC supercluster which also have member 
clusters of one or more other SCL superclusters. This implies that, in some cases for example,
the members of three SCL are redistributed in two MSCC. The statistics of the components,
fraction of measured redshift and fraction of isolated clusters for the
SCL, MSCC, and SSCC catalogues, are listed in Table~\ref{table:comparison}.

\begin{table}
  \centering
\caption{The first ten entries of the list of clusters used as input
catalogue for MSCC and SSCC. Columns are: (1) Abell/ACO number/ID (including
all line-of-sight components within $z\leq0.15$); (2) measured or estimated 
redshift;
(3) and (4) sequence number of the respective MSCC and SSCC supercluster host, 
or ``iso'' if the cluster is isolated in the respective catalogue, 
a ``$-$'' indicates that the cluster was not used in the respective input
catalogue (northern clusters for SSCC and S-clusters for MSCC); 
(5) in case of line-of-sight components, percentage of galaxies 
with measured redshift in each component is listed. The full table is available 
online.
}
\begin{tabular}{lcccc}
\hline
  \multicolumn{1}{c}{Abell/ACO} &
  \multicolumn{1}{c}{$z$}       &
  \multicolumn{1}{c}{MSCC}      &
  \multicolumn{1}{c}{SSCC}      &
  \multicolumn{1}{c}{\%}        \\
\hline
  A0001  & 0.1249 & 3    &  $-$  &  $-$ \\
  A0002  & 0.1225 & iso  &  iso  &  $-$ \\
  A0003  & 0.1022 & 4    &  $-$  &  $-$ \\
  A0005e & 0.1120 & iso  &  $-$  &  $-$ \\
  A0007  & 0.1030 & iso  &  $-$  &  $-$ \\
  A0008A & 0.1092 & 9    &  $-$  &  62  \\
  A0008B & 0.1441 & iso  &  $-$  &  38  \\
  A0012  & 0.1259 & iso  &  $-$  &  $-$ \\
  A0013  & 0.0946 & iso  &  iso  &  $-$ \\
  A0014  & 0.0653 & 1    &  10   &  $-$ \\
\hline\end{tabular}
\label{table:cluster}
\end{table}

A total of 114 (40\%) superclusters in SCL were labelled as candidates 
(``c''). Only for the sake of comparison, when we use the same criterion 
for ``candidate'' as \citet{eina01}, we find that 81 candidates in SCL 
are in fact not superclusters in MSCC (i.e.\ their members were not linked). 
Another two SCL superclusters were partially confirmed, i.e., they appear fragmented as two MSCC 
superclusters, one of them being still a candidate. Another 21 candidates in 
SCL are confirmed in MSCC, and a further 10 superclusters appear as candidates 
in both MSCC and SCL. In MSCC there are 37 (6\%) supercluster candidates 
(according to our definition), of which 25 are new superclusters, with 22 of 
these 25 exceeding the limit $z_{lim}=0.13$ imposed by \citet[]{eina01}.

\begin{table*}
  \centering
\caption{Brief comparison between SCL, MSCC, and SSCC. Columns are:
(1) catalogue acronym; (2) and (3) number of A/S clusters not considering line-of-sight 
components; (4) and (5) total number of A/S clusters including line-of-sight components; 
(6) and (7) fraction of A/S clusters with spectroscopic redshift in each 
catalogue not considering line-of-sight components (including line-of-sight components); 
and (8,9,10) fraction of input clusters that resulted to be isolated, members of pairs and 
members of supercluster ($m\geq3$) systems.} 
\begin{tabular}{lrrrrllccc}
\hline
                       &
  \multicolumn{2}{c}{Clusters} &
  \multicolumn{2}{c}{Components} &
  \multicolumn{2}{c}{Fraction of z$_{spec}$} &
  \multicolumn{3}{c}{Fraction of clusters that are} \\
     & A    & S    & ~~~A    & S~~~~   & ~~~~A   & S~~~~   &  Isolated & in Pairs & in Superclusters \\
\hline
SCL  & 1663 & $-$  & ~~~$-$  & $-$~~~  & 64.4\%        & $-$           &  30.1\%  & 15.5\% & 54.4\% \\
MSCC & 2531 & $-$  & ~~~3410 & $-$~~~  & 90.0 (92.6)\% & $-$           &  33.8\%  & 17.1\% & 49.1\% \\
SSCC & 974  & 869  & ~~~1217 & 1168~~~ & 87.4 (89.9)\% & 88.0 (91.1)\% &  36.5\%  & 18.7\% & 44.8\% \\
\hline
\end{tabular}
\label{table:comparison}
\end{table*}

\subsection{Notes on Individual Superclusters}
\label{sec:individual}
We searched the literature for previous studies of the richest 
superclusters in MSCC and SSCC (i.e.\ those 13 MSCC and 11 SSCC 
with $m\geq14$) in order to explore 
the reliability of these catalogues. Only one of these richest 
superclusters, SSCC~82, which exceeds $z=0.13$, had not been identified before.
This comparison also gave some clues on the differences between MSCC and 
SSCC caused by the inclusion of S-clusters in SSCC. 

\begin{figure*}
\centering\includegraphics[width=0.98\textwidth]{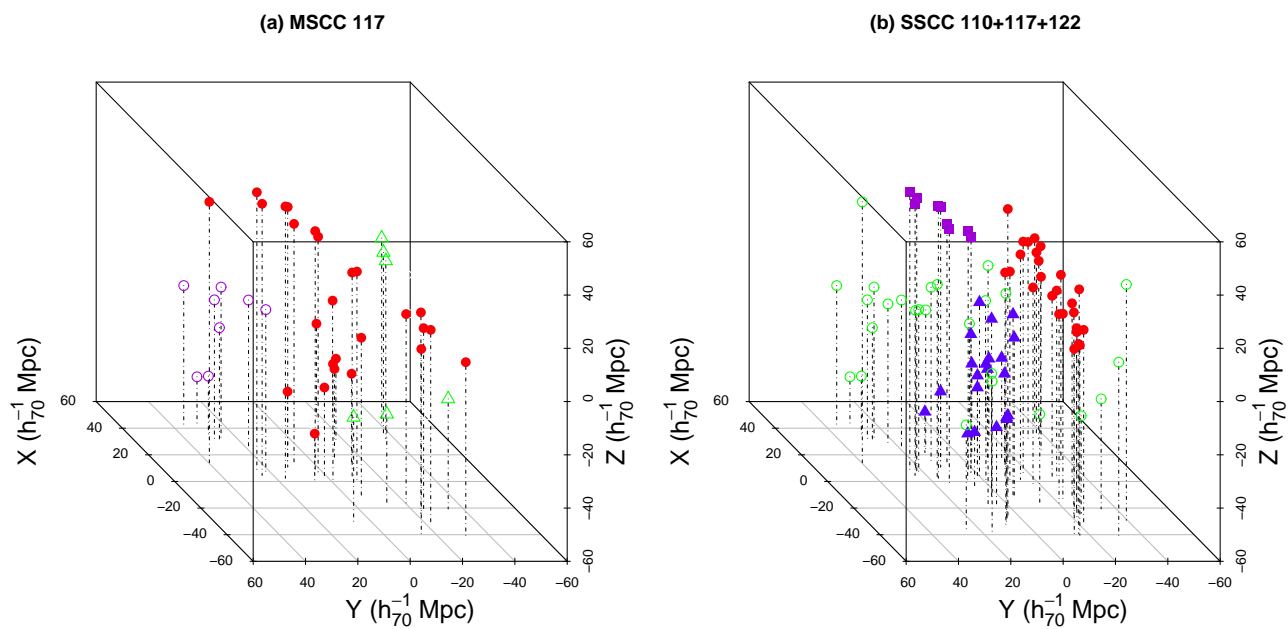}
\caption{{\it Horologium-Reticulum supercluster}. The XYZ coordinates
are derived from RA, DEC, and $z$ of the individual clusters, where 
X is pointing toward RA=DEC=0, and Z=0 is pointing to the north celestial pole. 
Left panel: distribution of clusters in the {\it Horologium-Reticulum} region 
according to the MSCC: 
solid circles are members of {\it Horologium-Reticulum} supercluster (MSCC~117); 
open circles are member clusters of MSCC~115, a foreground supercluster 
to {\it Horologium-Reticulum}; and open triangles are other clusters in 
the volume. Right panel: distribution according to SSCC; {\it 
Horologium-Reticulum} is fragmented into three superclusters: SSCC~110 
(solid circles), SSCC~117 (solid triangles) and SSCC~122 (solid squares);
open circles are other clusters in the volume.
}
\label{figure:hor}
\end{figure*}

\begin{figure*}
\centering
\includegraphics[width=0.98\textwidth]{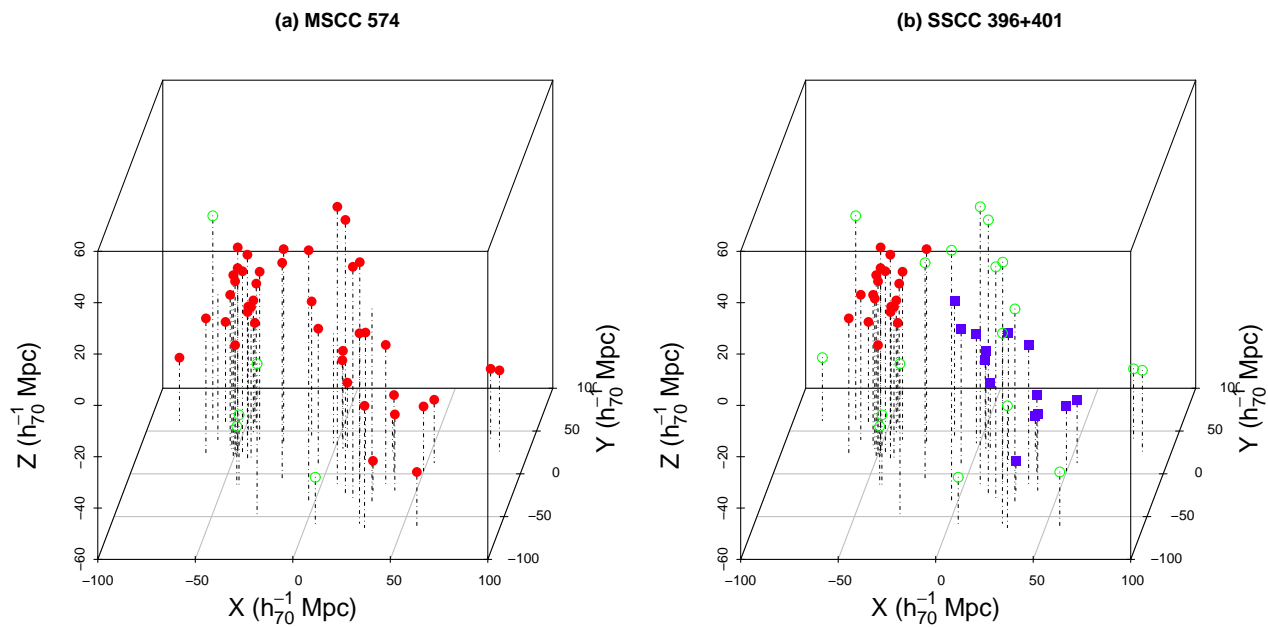}
\caption{{\it Aquarius~B} supercluster. XYZ coordinates as in Figure~\ref{figure:hor}.
Left panel: distribution of clusters in {\it Aquarius~B} according to MSCC: 
solid circles are members of {\it Aquarius~B}; open circles are 
other clusters in the considered volume. Right panel: distribution 
according to the SSCC; {\it Aquarius~B} is fragmented into two superclusters: 
solid circles are clusters in SSCC~401, taken as 
the main condensation of {\it Aquarius~B}; solid squares are members of 
SSCC~396, a filamentary structure toward higher redshifts;
open circles are other clusters in the considered volume.
}
\label{figure:aqub}
\end{figure*}

\begin{itemize}
\item[]\textbf{Sculptor (SCL\,009):} Identified as MSCC~33 and SSCC~36,
with similar multiplicities (24 and 22, respectively), although
more compact in the SSCC where 9 A-clusters were not linked while another
five S-clusters entered the supercluster. These S-clusters connect the main 
structure of MSCC~33 with the pair MSCC~47. The main member clusters,
A2798, A2801, A2804, A2811, and A2814, studied before by \citet{oba98}
and more recently by \citet{sato10}, remain in both catalogues as
the possible ``core'' of {\it Sculptor}.

\item[]\textbf{Pisces-Cetus (SCL\,010):} Is one of the 4 richest superclusters 
in SCL (together with Shapley, {\it Sculptor}, and {\it Aquarius}); fragmented
in 3 structures in our analisis: MSCC~39, the northern {\it Pisces-Cetus} (with
m\,=\,11); MSCC~27, the central {\it Pisces-Cetus} (m\,=\,9); and MSCC~1, the 
southern {\it Pisces-Cetus} (also m\,=\,9). The Southern {\it Pisces-Cetus}
is also found as SSCC 417, with m\,=\,14. It was studied previously by
\citet{porter2005} and \citet{porter2007}.

\item[]\textbf{Horologium-Reticulum (SCL\,048):} Studied by \citet{fleenor06}, 
{\it Horologium-Reticulum} is MSCC~117, the second richest in MSCC. 
Figure \ref{figure:hor} shows the distribution of its member clusters in 3-D. 
This supercluster appears significantly fragmented in SSCC, namely as SSCC~110, 
117, and 122.  This fragmentation is due to the smaller value of $\ell_c$
in the southern PF ($\sim15\,h_{70}^{-1}$ Mpc) compared to the 
master PF ($\sim22\,h_{70}^{-1}$ Mpc) at $z\sim0.06$ 
(cf.\ Fig.~\ref{figure:per}).

\item[]\textbf{Ursa Major (SCL\,109):} ~Is a northern supercluster listed
as MSCC~310. We found good agreement between the member clusters as listed 
in MSCC and those listed by \citet[][]{koko09}.

\item[]\textbf{Shapley (SCL\,124):} ~Is usually described
\citep[e.g.][]{bardelli94,proust06} as composed of three concentrations,
where the central concentration is the classical Shapley supercluster
\citep[integrated by A3556, A3558, A3560, A3564, and A3566, according to][]
{bardelli94}. This concentration was recognized by us as MSCC~389 and
SSCC~261. A second concentration, composed of A3528, A3530, and A3532,
is part of MSCC~389 and SSCC~249. The third condensation, called the 
``Front Eastern Wall'', with A3571, A3572, and A3575 as main clusters, 
was found by us as MSCC~401 (which contains other 7 clusters) and 
SSCC~267 (with other 5 clusters).

\item[]\textbf{Bo\"{o}tes (SCL\,138):} ~Identified as MSCC~414. 
In a recent analysis of superclusters in the SDSS region (using the member 
galaxies instead of the member clusters), \citet[][]{eina11a} separated it into
two superclusters (SCL~349 and SCL~351 according to their catalogue). These
superclusters were joined by us into MSCC~414.

\item[]\textbf{Corona Borealis (SCL\,158):} ~Is identified as MSCC~463. 
A recent analysis of \textit{Corona Borealis} by \citet{pearson14} suggests that 
many of the member clusters of MSCC~463 (namely A2056, A2061, A2065, A2067, 
A2089, and possibly A2092) form the only system, apart from the core of 
Shapley (MSCC~389 and SSCC~261), that shows conclusive evidence to be 
a bound supercluster.

\item[]\textbf{SSCC~82:} ~Is the richest supercluster in the SSCC, with
$m=38$. This supercluster appears in MSCC as three subsystems, namely 
MSCC~67 (A0210, A0214A, A2895, A2927B, and A2928C), 
MSCC~68 (A2923D, A2926B, A2927A, A2928B, A2931A, and A2932B), and MSCC~84 
with the remaining A-clusters of SSCC~82. These three superclusters are 
connected through member clusters S0193 and S0227A in the SSCC. 
\citet{eina01} included a fraction (nine clusters) of the clusters 
of SSCC~82 in SCL~232c.

\item[]\textbf{Aquarius A (SCL\,205):} MSCC~576 is known as 
{\it Aquarius A} \citep[][]{tully92}, or simply {\it Aquarius} \citep[][]{eina01}.
It corresponds to SSCC~402 and has
the same member clusters in both MSCC and SSCC. This membership is also in 
accordance with the identification of members made by \citet[][]{caretta02}. 

\item[]\textbf{Aquarius B (SCL\,209):} MSCC~574 is known as {\it Aquarius B} 
\citep[][]{tully92}. It is the richest system in the 
MSCC, with $m=42$. \citet[][]{batuski99} analysed the region of this 
supercluster, based only on A-clusters, and suggested it to be ``the 
largest supercluster in the Local Universe'' (in fact even connected to 
{\it Aquarius~A}). In our analysis it was separated into SSCC~401 ($m=20$) 
and SSCC~396 ($m=14$), with only a small aggregation of S-clusters to 
the systems. 
This separation is consistent with the study of \citet[][]{caretta02}, 
who considered both A- and S-clusters, also including poor clusters from 
other catalogs (namely APMCC, \citealt{dalton97}, and EDCC, \citealt{lumsden92}) 
and groups, and found that {\it Aquarius~B} is composed of two structures 
with only a low probability to be connected: a wall-like structure and 
a filament roughly along the line of sight towards higher redshift. 
Thus, the inclusion of S-clusters in our sample, the SSCC, reproduces this 
last result, as can be seen in the 3-D map of the member clusters of 
{\it Aquarius~B} in both MSCC and SSCC (Figure~\ref{figure:aqub}).

\item[]\textbf{Sloan Great Wall:} Another important structure is
the Sloan Great Wall \citep{gott05}.  \citet{eina11b} found this
structure to be integrated by the {\it Sextans} supercluster (SCL~88), {\it
Leo-Sextans supercluster} (SCL~91), {\it Virgo-Coma} supercluster (SCL~111), 
and SCL~126. We found the {\it Sextans} supercluster as MSCC~225,
while in our catalogue {\it Leo-Sextans} is fragmented into MSCC~247 and 273,
and {\it Virgo-Coma} into MSCC~311, 327, 343, and 352. 
SCL 126 was found by us as MSCC~376.

\end{itemize}

There are other 5 MSCC objects with $m\;\geq\;14$ that were not the subject of 
individual studies: MSCC~73 (SCL\,229), MSCC~76 (SCL~231), MSCC~84 and 94 (SCL~232c), 
and MSCC~123 (SCL~53, also known as {\it Fornax-Eridanus} supercluster). 
MSCC~238 has a multiplicity $m=21$, and its member clusters are 
line-of-sight components of clusters in SCL~256c, but they are not 
dominant components (they do not have significantly high fractions 
of galaxies with measured redshift along the respective observing
cone), except for A1028A. Another supercluster, MSCC~248, has most of the
dominant components of members of SCL~256c and was identified as its
equivalent. Another 3 superclusters with $m\;\geq\;14$ in SSCC have not been
studied in particular: SSCC~56 (SCL~22), SSCC~87 (SCL~232c), and
SSCC~300 (MSCC~505 with m\,=\,13; SCL~174, also known as {\it Microscopium} supercluster).

\section{Properties of the superclusters}
\label{sec:prop_sc}

\subsection{The effect of the inclusion of S-clusters}
\label{sec:matchsscc}

A comparison of the MSCC and SSCC catalogues should give us hints
on how the inclusion of S-clusters affects our description of the
nearby large-scale structure. We cross-matched them
to identify A-clusters in both catalogues, in order to see how the 
S-clusters affected their supercluster hosts.
The MSCC includes 211 (35\%) superclusters with $\delta\leq-17^\circ$.
Using the sample of A-clusters with $\delta<-17^\circ$, we found that only 132 
(62\%) of these 211 have an unambiguous match with only one supercluster in 
SSCC (the member clusters of one MSCC supercluster were distributed in only one
SSCC object). Another 29 (14\%) superclusters appear divided in two or more parts in 
SSCC, which is likely a consequence of the fact that the critical linking 
length, $\ell_c$, is smaller for the southern sample than for the Master~PF.
These fragments are not neccesarily less rich than their
MSCC counterparts, and often the inclusion of S-clusters as members 
maintains their multiplicities almost equal, but their physical sizes
are usually smaller, as pointed out in Section~\ref{sec:analysis}. 
This ``fragmentation'' is more dramatic in the richest superclusters ($m>15$).
As an example, in section~\ref{sec:individual} we observed a certain 
degree of fragmentation in the richest structures such as {\it Aquarius~B} 
and {\it Horologium-Reticulum}, into more compact or filamentary structures.  
Another 50 (24\%) MSCC superclusters disappeared in SSCC because their
member clusters were not connected by the smaller linking length.
Forty-seven A-clusters, members of some supercluster in MSCC, are only 
accompanied by S-clusters in SSCC. We shall refer to this effect
as ``nucleation''. 

We found that 290 isolated A-clusters of MSCC are also isolated in SSCC. 
Another 91 A-clusters that are isolated in MSCC became members of 
superclusters in SSCC. Of the total number of 423 superclusters in SSCC, 
194 are considered 
new superclusters, i.e.\ they do not have an unambiguous counterpart in MSCC 
nor do they arise from fragmentation. Of these, 83 (42\%) represent
nucleation of isolated clusters in MSCC,
3 (2\%) are bridges of S-clusters, i.e., isolated clusters in MSCC 
connected by S-clusters in SSCC, and 108 (56\%) are 
new superclusters formed by only S-clusters (including cluster pairs with
$m=2$). 

Although the SSCC covers only about 35\% of the whole sky (excluding the 
Galactic plane, $|b|\lesssim10^\circ$) the number of  superclusters
is about twice that in the MSCC in the same volume. As explained before, the main contributions 
to this enhancement of the number of superclusters are: 
nucleation (50\%, considering both, nucleation of disconnected 
A-clusters being members of MSCC superclusters, as well as nucleation of isolated 
A-clusters in the south), S-cluster systems (40\%), fragmentations (8\%), 
and S-cluster bridges (2\%).

\begin{figure}
\centering
\includegraphics[width=8cm]{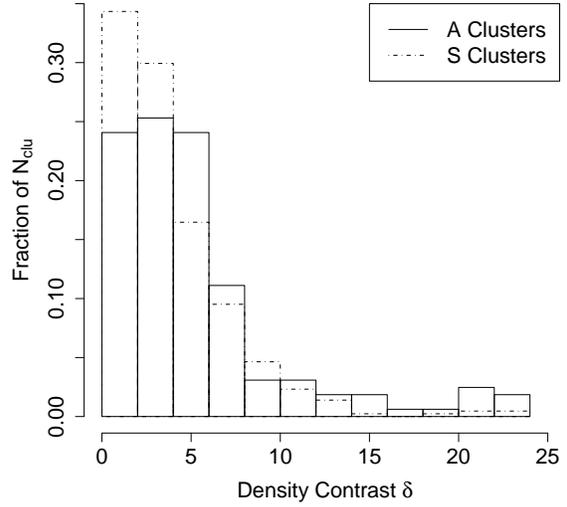}
\caption{Distribution of ambient density for Abell/ACO clusters (isolated+supercluster
members) in the southern sky ($\delta<-17^{\circ}$) for $z\leq0.06$, where the sample 
is reasonably complete for less massive systems. The density contrast is defined as the 
density we obtained for counting the number of any A- or S- cluster 
within a distance of 20\,$h^{-1}_{70}$ Mpc (the mean distance between 
clusters) around all clusters (including the cluster itself) 
divided by 2.8$\times10^{-5}$ $h^{3}_{70}$ Mpc$^{-3}$ (the mean cluster density of the
sample up to this redshift). Solid histogram: distribution for A-clusters. Dashed
histogram: distribution for S-clusters.
}
\label{figure:den}
\end{figure}

If S-clusters are poorer systems, it may be expected that they inhabit 
lower-density regions. To test this hypothesis, we determined the local 
number density of A- and S-clusters within 20\,$h_{70}^{-1}$\,Mpc
for each cluster in the southern sky ($\delta<-17^{\circ}$)
with redshift $z\leq0.06$. Out to this redshift, based on
an examination of Figure~\ref{figure:sel}, the sample seems to be
reasonably complete.  Figure~\ref{figure:den} shows the histograms for the local density
of southern clusters. It is apparent that S-clusters tend to inhabit lower-density 
regions than A-clusters.
Three statistical tests were applied to corroborate this visual inspection, 
the Kolmogorov-Smirnov (KS), Cram\'er and Tukey tests. The p-values
are listed in Table~\ref{table:sclu}, and suggest a low probability
that both distribution are drawn from the same population. This is in
agreement with the scenario where the main effect of S-cluster is the
nucleation, the fact that A-clusters are surrounded by S-clusters in
the external regions of superclusters, or in filaments, where the density is
lower. 

On the other hand, we did find 108 superclusters formed only by S-clusters,
but none of those has $m\;>\;4$, and indeed 71\% of these are pairs ($m\;=\;2$).

\begin{table}
  \centering
\caption{Statistical tests for density distributions between
A and S clusters showed in Figure \ref{figure:den}.
}
\begin{tabular}{cc}
\hline
Test   & p-value  \\
\hline
KS       & 0.01     \\
Cram\'er & $<$0.01  \\
Tukey    & $<$0.01  \\
\hline
\end{tabular}
\label{table:sclu}
\end{table}

\subsection{Multiplicity Functions}
\begin{figure*}
\centering
\includegraphics[width=0.98\textwidth]{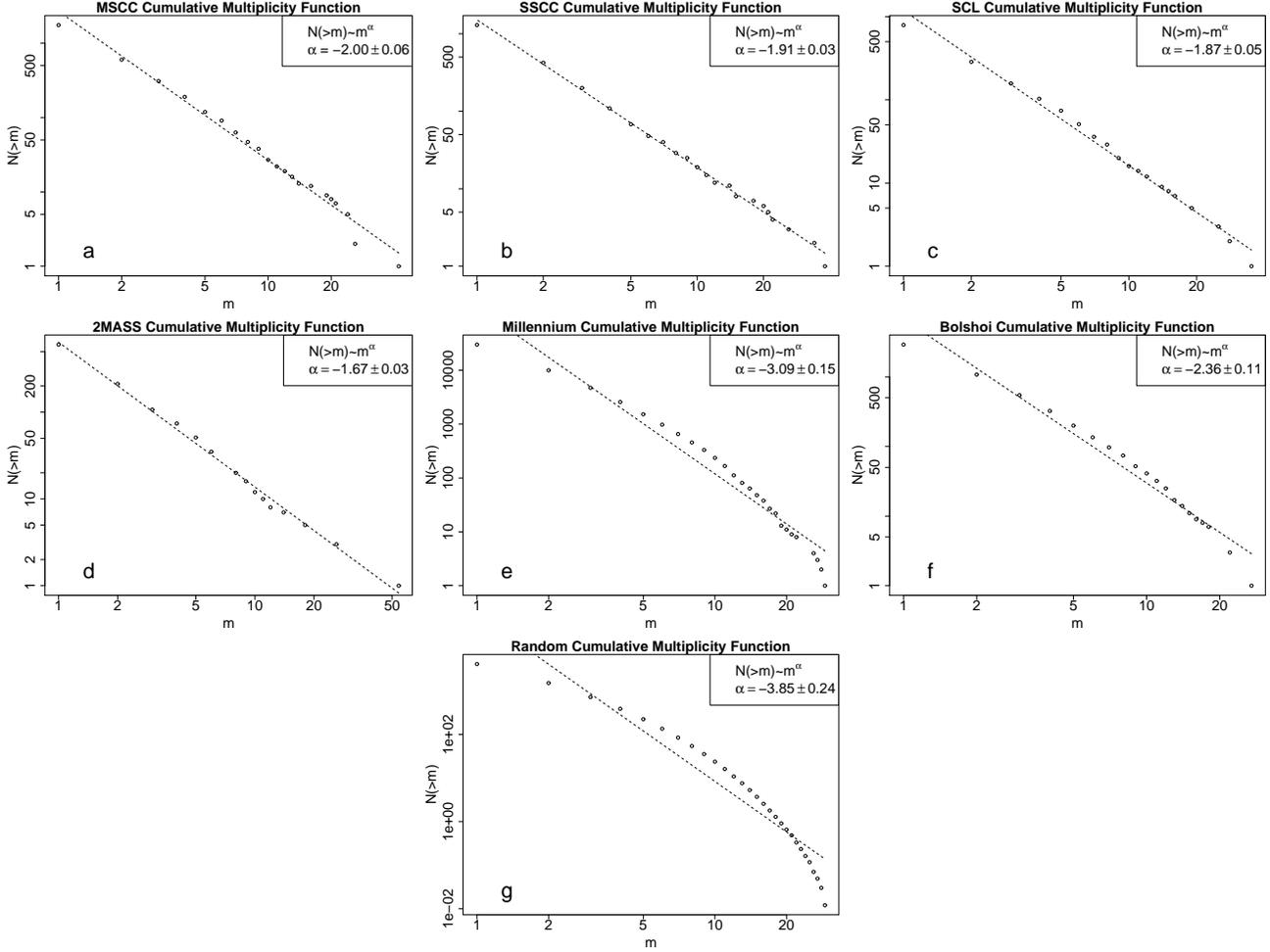}
\caption{Multiplicity functions for: (a) MSCC, (b) SSCC, (c) SCL, 
and supercluster catalogues based (d) on 2MASS groups,
(e) on a mock catalogue obtained from the Millennium simulation, 
(f) on a mock catalogue obtained from the BOLSHOI simulation, and 
(g) on the average of one thousand simulated mock catalogues of 
randomly distributed points with a mean density of 
$\bar{n}=7.4\times10^{-4}\;h^{3}_{70}$ Mpc$^{-3}$ (the cluster density of the 
input sample for MSCC). See the text for details.
}
\label{figure:mf}
\end{figure*}

The multiplicity function (MF) is the distribution of multiplicities
(the amount of member clusters) of the superclusters.
The cumulative MFs of MSCC and SSCC catalogues are shown in 
Figures~\ref{figure:mf}a and \ref{figure:mf}b. 
Both follow very closely a power law, for which we obtained the slope, 
$\alpha$, defined as $N(>m)\propto m^{\alpha}$, whose values are listed 
in each panel of that figure. 
We compared the MF obtained for MSCC and SSCC with the ones obtained for 
two other observational samples, as well as for two mock catalogues. 
We also compared them to one generated from a completly random sample.

The two other observational samples are the supercluster catalogue
of \citet{eina01}, the equivalent all-sky supercluster catalogue 
based on Abell/ACO clusters before our MSCC (Figure \ref{figure:mf}c), 
and the catalogue of 2MASS groups \citep[Figure \ref{figure:mf}d,][]{crook07}, 
to which we applied our FoF algorithm to create a comparison supercluster catalogue
including lower mass systems. \citet{eina01} show the MF for SCL, but these 
authors did not attempt a cumulative log-log plot of the multiplicity function 
which would show a power-law distribution. 
The SCL superclusters also follow a power law, though with a slightly 
higher dispersion than our MSCC and SSCC catalogues. The MFs of 
the three supercluster samples (MSCC, SSCC, and SCL) have similar slopes: 
$\alpha\,\sim\,-1.9$. Figure~\ref{figure:mf}d shows the MF for a 
supercluster catalogue based on 2MASS ``groups'' \citep{crook07}, which 
dominatly consists of systems of lower mass than the S-clusters 
in SSCC, but includes the few rich Abell clusters that exist within its 
redshift limit ($z_{lim}\sim0.033$). This catalogue, only produced to 
check the presence of a power law in its MF, was obtained 
using a PF similar to the ones generated for MSCC and SSCC in order 
to compensate for the selection effects with the redshift. 
The power law seems to apply to systems of very different mass,
and, since the SCL was produced using a single critical linking length, $\ell_c$,
for the entire volume, it also seems to be independent of the selection 
of $\ell_c$ to produce the catalogue.

Can the mock catalogues based on simulated data reproduce the MFs of the
observational samples? The distribution generated from mock catalogues
come from the Millennium simulation \citep[Figure \ref{figure:mf}e,][]{springel05}
and Bolshoi simulation \citep[Figure \ref{figure:mf}d,][]{klypin11}. 

The Millennium project is a $\Lambda$CDM N-body simulation of the evolution,
since redshift z=127, of $2160^3$ (cold dark matter) particles within a 
box of side 500\,Mpc based on $H_{0}\;=\;73$ km s$^{-1}$ Mpc$^{-1}$.
For the FoF calculations in the present paper all cosmological parameters
were scaled to $H_{0}\;=\;70\;h_{70}$ km s$^{-1}$ Mpc$^{-1}$, thus the
box side corresponds to $521\;h_{70}^{-1}$ Mpc. The mass of the individual
particles is M/M$_\odot=8.98\times10^8\;h_{70}^{-1}$.  
We selected all dark matter haloes obtained from the internal FoF of the 
Virgo-Millennium Database
(http://gavo.mpa-garching.mpg.de/Millennium/) with the condition
M/M$_\odot\geq10^{13}\;h_{70}^{-1}$ ($\geq11136$ particles per dark matter halo).
The dark matter halo sample obtained in this way has a density of similar order 
as that obtained for the sample of 2MASS groups (Table \ref{table:catbased}). 
We applied our own FoF to the 51998 massive dark matter haloes that follow 
the mass condition within the complete volume of the Millennium simulation, 
which corresponds to a redshift limit of $z\sim0.12$. 
Since the mean density is by construction the same throughout the Millennium volume,
we assumed a constant value of $\ell_c$ throughout the volume. 
Thus, $\ell_c$ was obtained considering  the total volume 
and not parts of it as MSCC and SSCC (and the sample of 2MASS group). 
However, it is clear from Figure \ref{figure:mf}e that the distribution of 
superclusters obtained from such haloes does not follow the power law 
obtained for observational samples.

The other mock catalogue (Figure~\ref{figure:mf}f) is based on the
sample of dark matter haloes of the Bolshoi simulation, a $\Lambda$CDM N-body 
simulation that uses $2048^3$ dark matter particles within a box of side
250 $h^{-1}_{70}$ Mpc. It has the largest resolution for a cosmological 
simulation of this type, not only in space but also in time evolution. For
comparison, Millennium is based on 11\,000 time steps per particle while 
Bolshoi was evolved since $z\;=\;80$ in $\sim400\,000$ time steps. 
The simulation was obtained with the
Adaptive-Mesh-Refinement (AMR)-type code \citep{kravtsov97} instead of the
GADGET code \citep{springel01} used for Millennium. For Bolshoi, we 
extracted from the MultiDark Database (http://www.multidark.org/MultiDark/)
such haloes with M/M$_\odot\geq10^{13}\;h_{70}^{-1}$ (haloes were determined by
an internal FoF of the MultiDark Project). The particle density of Bolshoi is,
again, comparable to that of the 2MASS group sample, and similar to the sample of
massive haloes of Millennium we used (cf.\ Table \ref{table:catbased}).
We applied our FoF to these haloes, using the same
procedure as for Millennium. The MF for superclusters we obtained in 
this way follows more closely the power law than the mock catalogue 
obtained with Millennium, but yet it is evident that neither of the 
two mock catalogues produce a cumulative MF that resembles the power-law 
distribution of multiplicities for the real superclusters.

\begin{table}
\centering
\caption{Comparison of properties of different samples of ``particles'' used
to study the multiplicity function. Columns are: (1) the name of the sample,
(2) the number of groups/clusters/particles within the considered volume,
(3) the volume in units of $h^{-3}_{70}$ Gpc$^3$, for MSCC, SSCC, SCL, and 
2MASS groups, excluding the zone of avoidance near the Galactic plane,
(4) the density, $\bar{n}$, of groups, clusters or dark matter haloes, 
excluding the zone of avoidance where neccesary, and (5) the slope $\alpha$ 
fitted to the cumulative multiplicity function.
}
\begin{tabular}{lrrrc}
\hline
\multicolumn{1}{c}{Catalogue}   & \multicolumn{1}{c}{N} & \multicolumn{1}{c}{Volume}  & \multicolumn{1}{c}{$\bar{n}$} & $\alpha$ \\
\multicolumn{1}{c}{based}       & {\tiny clu/par}       & ($h^{-3}_{70}$ Gpc$^3$)     & ($h^{3}_{70}$ Mpc$^{-3}$)     &          \\
\hline
MSCC         &  3410 & $82.5\times10^{-2}$ & $\leq7.4\times10^{-6}$ & $-2.00$ \\
SSCC         &  2385 & $30.5\times10^{-2}$ & $\leq2.8\times10^{-5}$ & $-1.91$ \\
SCL          &  1663 & $54.5\times10^{-2}$ & $3.0\times10^{-6}$     & $-1.87$ \\
2MASS groups &  1261 &  $1.0\times10^{-2}$ & $5.3\times10^{-4}$     & $-1.67$ \\
Millennium   & 51998 & $14.2\times10^{-2}$ & $3.5\times10^{-4}$     & $-3.09$ \\
Bolshoi      &  5585 &  $1.6\times10^{-2}$ & $3.4\times10^{-4}$     & $-2.36$ \\
Random       &  7329 & $99.8\times10^{-2}$ & $7.4\times10^{-6}$     & $-3.85$ \\
\hline
\end{tabular}
\label{table:catbased}
\end{table}

While the observational MF distributions for higher mass systems
are relatively well fitted by a power law with $\alpha\sim-2.0$,
the simulated ones do not follow closely a power law, presenting
a more curved distribution (lower fraction of superclusters with low and high
multiplicities, and larger for intermediate multiplicities than in a power
law regime), resulting in lower values for $\alpha$. 

What would one expect from a completely random sample of particles\,? 
Figure~\ref{figure:mf}g is the average of the MF of one thousand
simulated catalogues based on randonly distributed points that represent
Abell/ACO clusters. These distributions were produced assigning random
coordinates to each point. Each distribution contains 7\,329 points in a
spherical volume of ``radius'' $z=0.15$. This condition simulates the 
mean density of the best-sampled region of MSCC. In each sample, the 
value of the critical linking length, $\ell_c$, was found, and a FoF algorithm
was applied. We produced one thousand mock supercluster catalogues,
from which the number of frequency of each richness was averaged (note
that this resulted in high-richness systems having fractional values of
frequency in Figure~\ref{figure:mf}g). Again, as for the other distributions,
the slope $\alpha$ is shown, but neither the cosmological simulations nor the random
distribution reproduces the shape of the cumulative multiplicity function of the
observational samples.  Actually, the completely random sample appears 
more similar to the distribution based on the Millennium data.

The hump in the curves based on mock catalogues, much stronger in the
random and Millennium MF, can be interpreted as a lower probability to
obtain high-richness superclusters from such samples. We can see this in the
random distribution, where the structures are in a Poissonian regime and the
value of the power spectrum is expected to be zero. However, in a
gravitational regime one may expect to obtain structures (power spectrum $>$\,0), 
that drive the superclusters toward the richest systems, which would explain
why in the MF of cluster-based catalogues the curves do not decay (at the rich end)
as abruptly as in the random case. 

Power laws have also been fitted in the literature for the number of galaxies 
within groups of SDSS by \citet{berlind06} and
\citet{nurmi13}. \citet{berlind06} found values for $\alpha$ of $2.49\pm0.28$,
$2.48\pm0.14$ and $2.72\pm0.16$ for different volume-limited samples
of absolute magnitude limits of $M_r\leq-18$, and $-19$, $-20$. On the
other hand, \citet{nurmi13} found $2.02\pm0.18$, $2.12\pm0.17$, $2.26\pm0.19$
for these luminosity ranges, and $3.29\pm0.24$ for $M_r\leq-21$. Since these
distributions are based on ``particles'' (galaxies) of much lower mass we do not
expect a direct relation with those obtained here, except, as we
said, that the power law will be a natural distribution for systems in
gravitational regimes.

\section{Conclusions and Summary}
\label{sec:conclusion}
We summarize our conclusions as follows.

\begin{itemize}

\item[1.] We constructed two new supercluster catalogues based on the
optically-selected Abell/ACO clusters, which not only reach deeper in space
($z\;\leq\;0.15$) than previous ones, but also contain a much higher fraction
($\gtrsim$\,85\%) of clusters with spectroscopically confirmed redshifts.
Different from previous works, we include in our analysis the different 
line-of-sight components for the clusters (up to the above-mentioned 
redshift limit). One of these catalogues is the all-sky Main SuperCluster 
Catalogue (MSCC), based on only the rich Abell A-clusters, the other
is the Southern SuperCluster Catalogue (SSCC) covering declinations 
$\delta\;<\;-17^\circ$ and including the supplementary Abell S-clusters.

\item[2.] A tunable linking length was used to generate these supercluster
catalogues, taking into account the undersampling at higher redshift and
the different selection functions in different regions of the sky.
The area covered by SDSS is clearly better sampled than the rest of the 
sky, and a special percolation function was fitted for this region. 
Also, for the southern sky, the inclusion of the S-clusters required
an independent correction for the selection function.
We conclude that the ``percolation function'' is an efficient tool for
detecting and correcting the selection function for samples of galaxy
clusters.

\item[3.] For the all-sky main sample of A-clusters we found that the
maximum completeness is reached near $z\;\sim\;0.06$, which leads to
a critical linking length of $\ell_c\sim 22.5\,h_{70}^{-1}$ Mpc, while for
the south ($\delta\;<\;-17^\circ$) the S-clusters provide the best
sampling at $z\,\sim\,0.01$, leading to $\ell_c\sim11.5\,h_{70}^{-1}$ Mpc.

\item[4.] The MSCC and the SSCC contain 601 and 423 superclusters, 
respectively.  Considering the probability for a cluster to be member
of a supercluster or not, we found the following: 35\% of the clusters
tend to be isolated, 18\% form pairs and 47\% are hosted by a supercluster
with $m\ge3$.
Of the superclusters found, the fraction of pairs ($m\,=\,2$) is $\sim50$\%,
while the fraction of very rich superclusters ($m\,>\,10$) is $\sim4$\% of the
systems. The complete catalogues and input lists of clusters are available 
in the electronic version of this paper.

\item[5.] By comparing our MSCC directly with the one by \citet{eina01}
we find: (1) a slightly higher fraction of isolated clusters (35\% compared to 
30\% in SCL); (2) 37 new superclusters identified in the same volume 
investigated previously; (3) 126 new superclusters near and beyond $z\;=\;0.13$; 
(4) 70\% of Einasto et al.'s candidate superclusters were not confirmed as such. 
The fraction of candidate superclusters was reduced from 40\% to 6\%, 
thanks to the much higher fraction of clusters with spectroscopically
confirmed redshifts in our sample.

\item[6.] Comparison between the MSCC and SSCC reveals that the S-clusters
seem to prefer to inhabit the surroundings of richest clusters.  The lower 
critical linking length in SSCC only breaks up the MSCC systems with 
$m\;\geq\;15$ and, since these systems represent only $\sim$2.0\% of the MSCC, 
and only 1.2\% of them are in the declination range of SSCC, this 
fragmentations is not significant. More important effects of the inclusion of
S-clusters on the distribution are new superclusters (with multiplicities 
$m\;\geq\;2$) of only S-clusters and the ``nucleation'' of A-clusters
surrounded by S-clusters.

\item[7.] The distributions of supercluster ``richness'' (the multiplicity 
function) of the MSCC and SSCC follow a power law of slope $\alpha\sim-2.0$. 
A very similar behaviour is seen in other catalogues based on observational 
data, namely the SCL \citep{eina01} and a catalogue generated from 2MASS 
groups \citep[][]{crook07}, the latter being far more complete for lower-mass
systems. However, when similar supercluster catalogs are generated from mock
data, namely using the Millennium \citep{springel05} and the Bolshoi
\citep{klypin11} N-body cosmological simulations, the multiplicity functions 
change from a power law to a convex curve, predicting less superclusters
in both the low- and high-multiplicity regimes. A similar behaviour, even
farther from a power law, is found for an entirely random distribution 
of input clusters.

\end{itemize}

\section*{Acknowledgments}
We thank Kristin Riebe and the people of the MultiDark Project as well as
Gerard Lemson and the people of the Virgo Consortium for the access 
to the databases of the cosmological simulations used in this paper.
We also thank the \citet{rproject} and other developers of R Project libraries:
\citet{liggesmachler2003}, \citet{franz2006}, \citet{hbw2008},
\citet{warnes2014}, and \citet{zeileis2004}. All graphics and
statistics in this work were generated using these tools.
M.C.-M. acknowledges a CONACyT grant for a Master thesis, and C.C. and H.A.
a Guanajuato University grant \#219/13. C.C. and H.A. are also grateful 
to the staffs of the following observatories, where some of the spectroscopic 
data used in this paper were obtained: Observatorio Astron\'omico Nacional 
(Ensenada, Mexico), Observatorio do Pico dos Dias (Bras\'opolis, Brasil), 
South African Astronomical Observatory (Cape Town, South Africa) and 
Indian Astronomical Observatory (Hanle, India).


\end{document}